\documentclass{emulateapj}

\newcommand{\myemail}{nicolas.iro@nasa.gov}
\newcommand{\drakeemail}{leo.d.deming@nasa.gov}

\usepackage{color}

\usepackage{aas_macros} %%% includeADS acronyms in the biblio

\newcommand{\rsol}{R$_\odot$}
\newcommand{\rj}{R$_{\rm J}$}
\newcommand{\mj}{M$_{\rm J}$}

\shorttitle{Atmospheric model for eccentric transiting exoplanets}
\shortauthors{Iro \& Deming}

\begin{document}

\title{A time-dependent radiative model for the atmosphere of the eccentric
       exoplanets}

\author{N. Iro\altaffilmark{a} and L. D. Deming}
\affil{NASA/Goddard Space Flight Center, Planetary Systems Laboratory,
              Code 693, Greenbelt, MD 20771,USA; \myemail; \drakeemail}
%\email{\myemail}
%\email{\drakeemail}

\altaffiltext{a}{LESIA, Observatoire de Paris-Meudon, place Jules Janssen,
              92395 Meudon Cedex, FRANCE}

\begin{abstract}
We present a time-dependent radiative model for the atmosphere of
 extrasolar planets that takes into account the eccentricity of their
 orbit.  In addition to the modulation of stellar irradiation by the
 varying planet-star distance, the pseudo-synchronous rotation of the
 planets may play a significant role.  We include both of these
 time-dependent effects when modeling the planetary thermal structure.
 We investigate the thermal structure, 
and spectral characteristics for time-dependent stellar heating
for two highly eccentric planets.
 Finally, we discuss observational aspects for those planets suitable
 for \emph{Spitzer} measurements, and investigate the role of the
 rotation rate.
\end{abstract}


\section{Introduction}

Observations of close-in extrasolar planets using the\emph{Spitzer Space
Telescope} have permitted the inference of their atmospheric properties
via detection of their emergent infrared light
\citep{Deming07,Knutson07,Grillmair07,Swain08}. 
Most
investigations of this type are not sensitive to the time variation of
the planet's properties.  However, a real planet will sometimes
produce a dramatically time-variable infrared signal in response to
several variables, as was recently demonstrated by \citet{LL08}.
For planets on eccentric orbits, the observed signal will vary due to
the star-planet distance, as well as rapid changes in our viewing
angle to the planet.  The latter is sensitive to the
pseudo-synchronous rotation rate near periastron for planets on
eccentric orbits.  Also, there is the finite radiative response time,
wherein the planet's atmosphere does not heat or cool instantaneously
in response to variable stellar irradiation. Finally, a vigorous
hydrodynamics is expected to re-distribute the heating by stellar
irradiation.

A full treatment of time-dependent radiative transfer coupled to
nonlinear hydrodynamics is beyond the current state of the art.  In
this paper, we concentrate on the effect of time-dependent radiative
transfer in isolation from the re-distribution of heat by dynamics.
There are two motivations for investigating time-dependent radiative
effects in relative isolation, coupled only to the changing rotational
aspect geometry.  First, we want to know the time scales for the
exoplanet atmosphere to respond as a function of orbit eccentricity
and depth in the atmosphere.  Second, a comparison of our calculations
to \emph{Spitzer} time series data for eccentric planets might in principle
inform us of the degree to which the observations are affected by
dynamics, i.e., can simple time-dependent radiative transfer and
changing aspect geometry account for most of the observations, or do
hydrodynamic effects become dominant on a short time-scale?
Planets on eccentric orbits receive a highly varying flux from their stars.
Even in the case of 
one of the mildest eccentric planets WASP-10 b, with an
eccentricity of 0.057 \citep{Christian08} this represents a variation
$\sim$~25\% in the received flux between periastron and apastron.
Our model takes into account the time-variable stellar irradiance impinging
on the atmosphere of these planets.

In Section~\ref{model}, we present our time-dependent model calculation
and assumptions.
In Section~\ref{planets}, we will point out some aspects of the eccentric
exoplanets.
We will focus our study on the 
 two transiting planets that have the largest eccentricities:
HD~80606b ($e = 0.932$) and HD~17156b ($e=0.67$).
A summary of the results and a conclusion will be presented in
Section~\ref{discuss}.

\section{The atmospheric model}
\label{model}

\subsection{Radiative transfer} \label{transrad}

The energy equation relating the evolution of the temperature profile in 
an atmosphere under hydrostatic equilibrium is:
\begin{equation}\label{eq:dTdt}
\frac{d T}{d t}= h(p) - c(p) {\rm,}
\end{equation}
where $h(p)$ is the heating rate and $c(p)$ the cooling rate.
Assuming that the energy flux is purely radiative, we have
$h(p) = -\frac{m g}{C_{\rm p}}\frac{d F_\star}{d p}$
and
$c(p) = -\frac{m g}{C_{\rm p}}\frac{d F_{\rm IR}}{d p}$.
We divided the net flux into the thermal flux emitted by the atmosphere 
$F_{\rm IR}$ (upward $-$ downward) and the net stellar flux $F_\star$ 
(downward $-$ upward), so that $F=F_{\rm IR} - F_\star$. 
The numerical method used to solve the energy equation 
as well as the sources of opacities
is discussed in detail in \citet{Iro05}.
 Once the radiative solution is reached,
the (temperature-dependent) adiabatic lapse rate is calculated for each layer
and it is imposed to the final solution where it would be exceeded.

\subsection{Input parameters}

Table~\ref{tab:input_param} summarizes the input parameters pertaining
to each planet, star and orbit among the eccentric transiting
planets.  An exception would be
the metallicity, and we adopted a solar metallicity for all the models
in this study.  In this table, we included HD~209458b, even though its
eccentricity is consistent with a circular orbit.

\begin{deluxetable*}{l|c|c|c|c|c|c|c|c|c|c|c|c|c}
\tabletypesize{\scriptsize}
\tablecaption{Planetary and stellar input parameters.\label{tab:input_param}}
\tablewidth{0pt}
\tablehead{&\multicolumn{6}{c|}{{Orbit}}&\multicolumn{3}{c|}{{Planet}}&\multicolumn{3}{c|}{{Star}}&\multicolumn{1}{c}{{Refs.}}}
\startdata
System         & $e$  & $a$  & $d_{\rm min}$ & $d_{\rm max}$ & $P$  & $\omega$ & $R_{\rm pl.}$  & Mass  & $P_{\rm spin}$  & Type    & $R_{\star}$ &$\left[\frac {\rm Fe}{\rm H}\right]$& \\
         &      & [AU] &  [AU]         &      [AU]     & [d.] &[$^\circ$]& [\rj]          &  [\mj]& [d.]            &         & [\rsol]     &      &\\

\tableline
HD~80606 &0.93 &0.432\tablenotemark{a} &0.029         &0.835          &111.45&300       & 1.1\tablenotemark{a}            & 4.18\tablenotemark{a}  & 1.72            & G5      & 1.0        &  0.43 & (1)\\
HD~17156 &0.67  & 0.159 & 0.052 & 0.266 & 21.22 & 121 & 1.23 & 3.09 & 3.80 & G0V & 1.47 & 0.24 & (2;3)\\
\(\left\{{{\rm HAT-P-2} \atop {\rm HD147506}}\right. \)&0.52 & 0.068 & 0.033 & 0.103 & 5.63 & 190 & 0.95 & 8.62 & 1.87 & F8 & 1.42 & 0.11 & (4)\\
XO-3 & 0.260 & 0.046 & 0.034 & 0.058 & 3.19 & 346 & 1.22 & 11.79 & 2.25 & F5V & 1.38 & -0.18 & (5)\\
HAT-P-11& 0.198& 0.053 &0.043  & 0.063 & 4.89 & 355 & 0.42 & 0.081 & 3.94 & K4 & 0.75 & 0.31 & (6)\\
GJ~436 & 0.15 & 0.085 & 0.024 & 0.033 & 2.64 & 351 & 0.39 & 0.070 & 2.32 & M2.5 & 0.47 & -0.32 & (7-10)\\
WASP-14 &0.095 & 0.037 & 0.033 & 0.041 & 2.24 & 255 & 1.26 & 7.725 & 2.12 & F5 & 1.297 & 0.0\tablenotemark{b} & (11)\\
HAT-P-1&0.067\tablenotemark{c} & 0.055 & 0.052 & 0.059 & 3.09 & 0\tablenotemark{d} & 1.225 & 0.524 & 3.01& G0V & 1.135 & 0.13 & (12)\\
WASP-10 & 0.057& 0.037 & 0.035 & 0.039 & 3.52 & 157 & 1.29 & 3.06 & 3.45 & K5& 0.784 & 0.03\tablenotemark{b} & (13)\\
WASP-6 & 0.054&0.042 &0.040 &0.044 & 3.36 & 99 & 1.22 & 0.50 & 3.30&G0V & 0.87 & -0.2 &(14)\\
WASP-12 & 0.049&0.023 &0.022 & 0.024& 1.09 & 286 & 1.79 & 1.41 & 1.07& F& 1.57 & 0.3\tablenotemark{b}&(15)\\
HD~209458 &0.014\tablenotemark{e}& 0.045 & 0.044 & 0.046 & 3.52 & 80\tablenotemark{e} & 1.31 & 0.69 & 3.52 & G0V & 1.46 & 0.01&(16)\\
\tableline
\enddata
\tablecomments{The metallicity is shown only for informational purposes since we used a solar metallicity for our models.The minimal and maximal distances (resp. $d_{\rm min}$ and $d_{\rm max}$) as well as the rotation period given by Eq.~\ref{eq:Hut81} ($P_{\rm spin}$)  are calculated as a function of the parameters found in the references.}
\tablenotetext{a}{\citet{Moutou09} analyzed the transit of HD~80606 and found slightly different values with little consequences on our study.}
\tablenotetext{b}{In fact, it is the {\it global} metallicity [M/H].}
\tablenotetext{c}{This value is an upper limit but will serve as guideline in this study.} 
\tablenotetext{d}{\citet{Winn08} measurements were also consistent with a zero eccentricity.}
\tablenotetext{e}{The measurement by \citet{Laughlin05} are also consistent with $e=0$. In that case, the argument of the periastron would be irrelevant.}
\tablerefs{
(1) \citet{LL08} ;
(2) \citet{Gillon08} ; (3) \citet{Fischer07}
(4) \citet{Loeillet08} ; 
(5) \citet{Winn08} ; 
(6) \citet{Bakos09}
(7) \citet{Deming07} ; (8) \citet{Demory07} ; (9) \citet{Bean06} ; (10) \citet{Maness07} % GJ436
(11) \citet{Joshi08} ;
(12) \citet{Johnson08} ; % HAT-P-1b
(13) \citet{Christian08} ;
(14) \citet{Gillon09} ;
(15) \citet{Hebb08} ;
(16) \citet{Laughlin05}.
}
\end{deluxetable*}

\subsection{Time-dependent calculations}

Since we use a time-marching algorithm
to solve for Equation~\ref{eq:dTdt}, we are able to include the variation
of the incoming insolation.
This variation is due to the non constant distance of the planet
from its star during the eccentric orbit, as well as the rotation of the
planet due to the pseudo-synchronous sate.
This incoming flux can be written as a function of time as follows:
\begin{equation}\label{eq:Fstar}
F_\star^{\rm inc}(t,\lambda)= \left( \frac{R_\star}{d(t)}\right)^2
\cos \left[\theta(t)\right] F_\star^0(\lambda) {\rm,}
\end{equation}
where $F_\star^{\rm inc}$ is the wavelength-dependent incident stellar flux at
the top of the atmospheric model, $R_\star$ is the stellar radius, $d(t)$ is
the time-dependent star-to-planet distance, $\theta$ is the planetary longitude
of the point of the atmosphere that is substellar at $t=0$ and
$ F_\star^0(\lambda)$ is the flux
emerging from the star.
At each time step, we calculate the planet--star distance ($d(t)$) as well as
the rotation ($\theta(t)$), and the subsequent
incident stellar flux.

A current limitation of our method is that the heating rates and cooling rates
as well as the abundance profiles of the atmospheric constituents are
calculated only once for the initial conditions.
We calculate them at the mean distance
(semimajor axis) under the condition where the incoming stellar flux
is averaged over the planet. However, the influence of the initial
thermal profile chosen as well as the heating and cooling rates
on the final results is only minor, as we will discuss
in Section~\ref{discuss}.

\section{Study of the planets}
\label{planets}

\subsection{Eccentric orbit}

\begin{figure*}[t!]
  \begin{center}
    \includegraphics[angle=90,width=.85\textwidth]{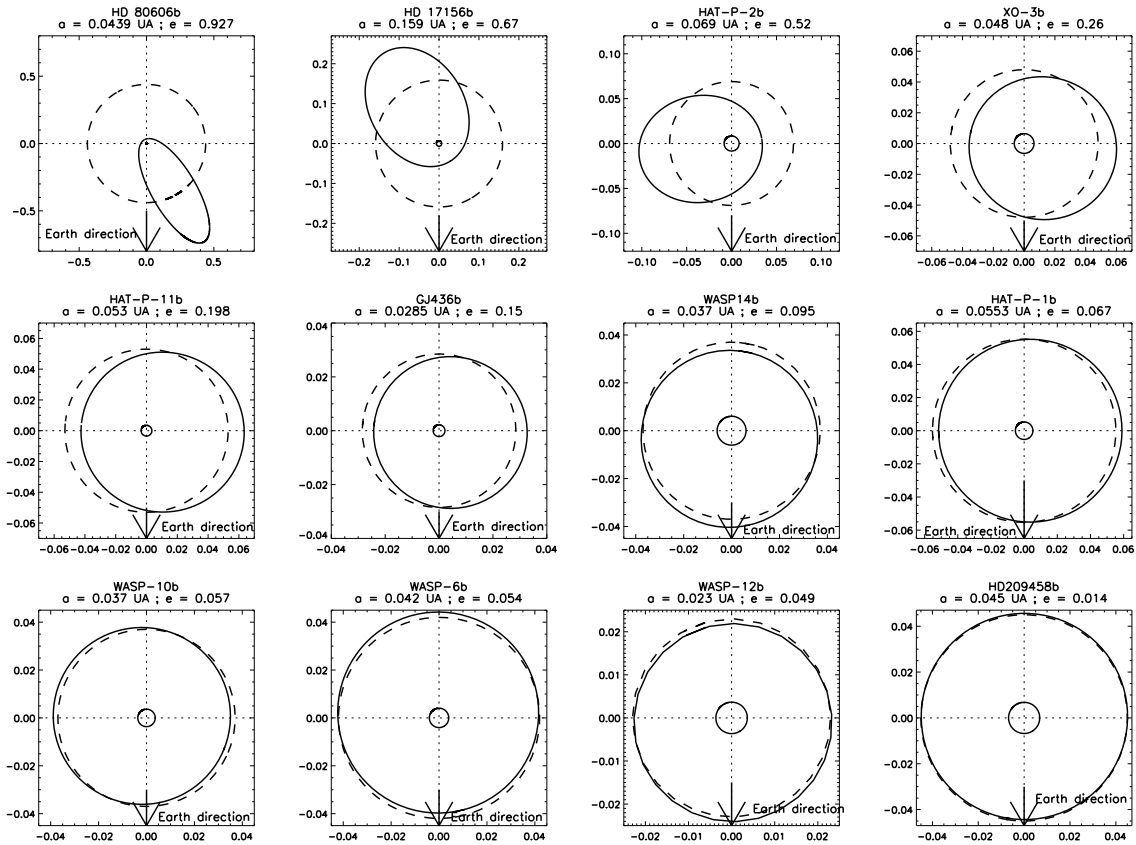}
  \end{center}
  \caption{Representation of the eccentric transiting planets known so far (circa February 2009)
    with parameters taken from Table~\ref{tab:input_param}. 
    The center circle symbolizing the star is to scale with respect to the orbit.
    The arrows indicate the direction to the Earth.
    The dashed circles show the semimajor axis distance (orbit without eccentricity).
    Note the different scales for each planetary system.\label{fig:orbits}
}
\end{figure*}

Figure~\ref{fig:orbits} shows the orbit for each eccentric transiting planets 
with the parameters taken from
Table~\ref{tab:input_param}.

Another effect that must be included is that an eccentric planet is
expected to be in pseudo-synchronous rotation due to the strong tidal
forces applied to the planet during the passage to periastron.
\citet{Hut81} calculated the remnant spin period as follows:
\begin{equation}
\label{eq:Hut81}
P_{\rm spin} =
\frac{\left( 1+3e^2+ \frac{3}{8}e^4\right) \left( 1-e^2\right)^{3/2}}
{1+\frac{15}{2}e^2+\frac{45}{8}e^4+\frac{5}{16}e^6}
P_{\rm orbit} {\rm .}
\end{equation}
The numerical value of the pseudo-synchronous rotation period
of the eccentric transiting planets is listed in Table~\ref{tab:input_param}.

\subsection{Individual planets}

\subsubsection{HD~17156b}

Discovered by \citet{Barbieri07}, this long period (21.2~days) planet has the
 \emph{second} largest eccentricity (0.67) among the transiting planets.

\paragraph{Previous studies}

\citet{Irwin08} performed a photometric analysis of the system which gave
approximately the same parameters as \citet{Gillon08} summarized in
Table~\ref{tab:input_param}
(except for a slightly larger planetary radius)
and found no transit timing variation due to the presence of additional planets
in the system.

They also presented a hydrodynamic model for this planet that predicted theoretical light curves for \emph{Spitzer}
observations calculated by using the climate model of \citet{LL07} applied to
the HD~17156b parameters.
They predicted the planet to star ratio $F_{\rm p}/F_\star$ in the 8 $\mu$m
band to vary from
$\sim 1.7 \times 10^{-4}$
to $\sim 5 \times 10^{-4}$
in the $\sim 30$ hr following periastron.
The authors used the \citet{Hut81} formula to obtain $P_{\rm spin} \sim 3.8$ days
with HD~17156b parameters.

\paragraph{Thermal structure}

\begin{figure*}
%  \epsscale{0.5}
\begin{center}
  \includegraphics[width=0.8\textwidth]{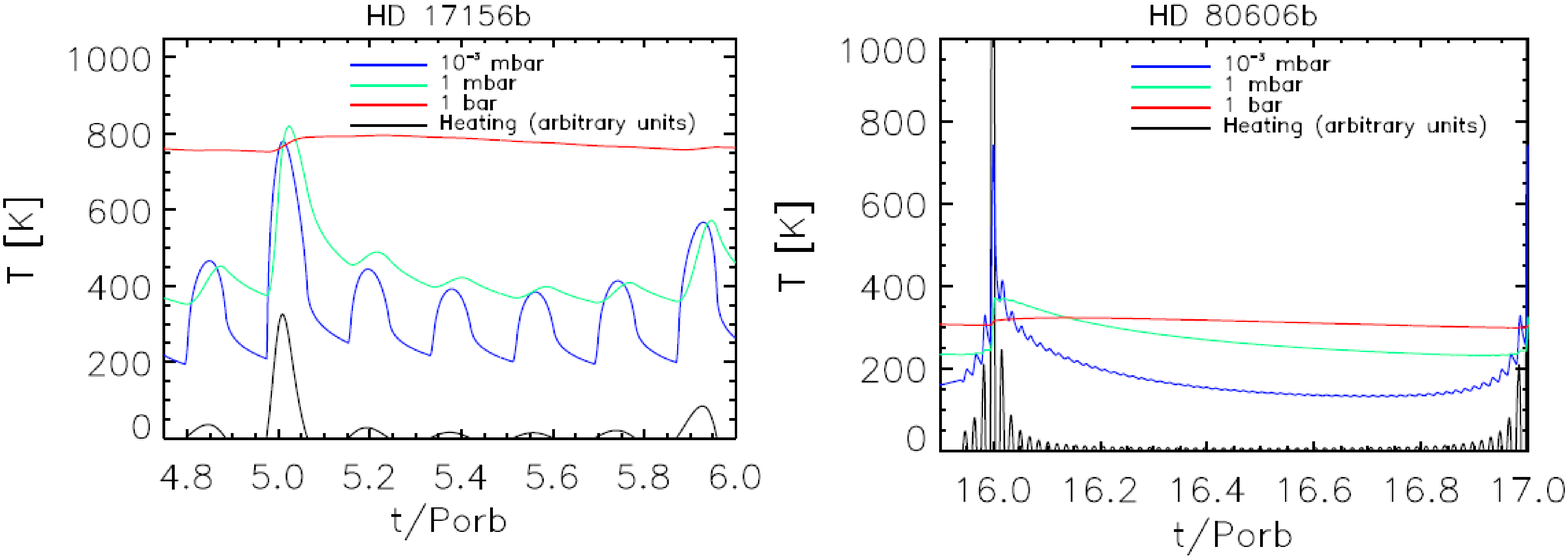}
\end{center}
  \caption{ Temperature of the planet atmospheres, for a single longitude at selected pressure levels 
 (10$^{-3}$~mbar, 1~mbar and 1~bar)
as a function of time\label{fig:res1}.
 The heating is a factor that is applied to the incident flux with respect to an averaged incoming flux model ($F_\star^{\rm inc}/F_\star^0$),
as defined by Equation~\ref{eq:Fstar}.
    We let the calculation make several orbital revolution in order to reach a periodic solution (typically 5 for HD17156b and 16 for HD80606b).
Note the different scales in temperature between the two planets.
}
\end{figure*}

In Figure~\ref{fig:res1}, we can show the temperature of one latitude during an
orbit for several pressure levels;
note the $\sim 5.6$ rotations during one orbit.
The low pressure levels are very sensitive to the heating variations
(the temperature variation of an atmospheric parcel is more
than 500~K at 1~mbar).
Deeper than 1~bar, the temperature variations are smaller than 50~K.
As previously discussed in \citet{Iro05}, for pressure greater than 1~kbar,
the thermal structure is mainly driven by the internal heat flux
originating from the planet interior, constrained by planetary evolution models.
Moreover, due to the longer radiative timescale at larger pressure,
there is a delay between the heating and the temperature increases.

\begin{figure*}
\begin{center}
  \includegraphics[width=0.8\textwidth]{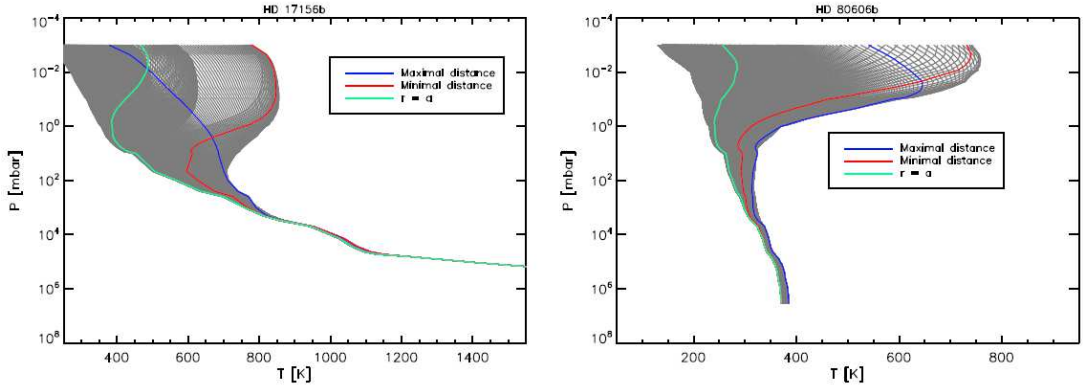}  
\end{center}
  \caption{ Temperature structures of the atmospheres for several orbital position ({\it lines})
    and for each time steps ({\it shaded area}). Note the different scales.\label{fig:tempe}
}
\end{figure*}

Figure~\ref{fig:tempe} shows all the thermal structures of one latitude during
an orbit.
We can see again that deeper than 5~bar, the atmosphere is only weakly
dependent
 on the incoming heating variation.
On the other hand, above the 100~mbar level, some of the profiles exhibit
a temperature inversion.
This inversion is generated by the strong heating by the star when the planet
is closer---near periastron.
It should be noted that no change of chemistry is taken into account here so the change of irradiation is sufficient for causing this inversion.
The inversion reaches an amplitude
of $\sim$~300~K.
This inversion has been previously inferred for highly irradiated planets
\citep{Burrows06, Fortney08}.
\citet{Fortney08} distinguished two types of planets by the degree of incoming
radiation causing or not this inversion
(pM class planets experiencing a very high stellar radiation exhibit
an inversion whereas cooler pL class planets do not).
As we show here, the highly eccentric planets are then likely to transfer from one class to the other during the course of their orbit.

\begin{figure*}
\begin{center}
  \includegraphics[width=0.8\textwidth]{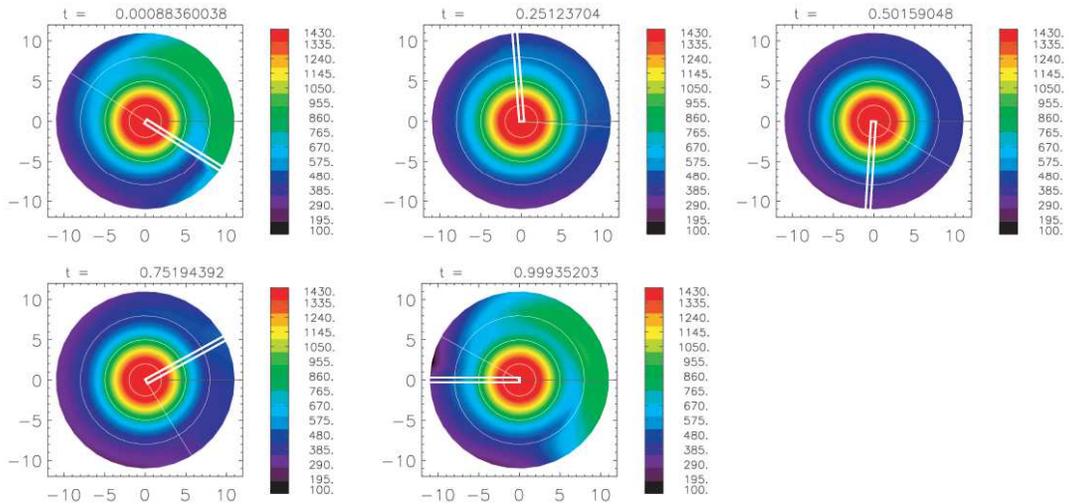}
\end{center}
  \caption{Thermal structure of the planet HD17156b atmosphere for selected times\label{fig:res_movie} during the orbit ($t/P_{\rm orb}$ = 0; 0.25; 0.5; 0.75 and 1).
    The axis represent the pressure in logarithmic scale. 
    The circles from center out indicate the 10$^{3}$, 1 and 10$^{-3}$ bar levels.
    The substellar point direction is indicated by the black line,  fixed at a single location to facilitate comparison between frames.
    The white line indicates the direction of the Earth.
    The time is shown as fractions of one orbit. 
%    This figure is also available as an mpeg animation in the electronic edition of the journal.
 The area contained in the rectangle is the parcel of atmosphere whose temperature evolution is plotted in Figure~\ref{fig:res1} (in pseudo-synchronous rotation).\\
An animation of this figure is available in the online journal.
  }
\end{figure*}

Figure~\ref{fig:res_movie} represents an equatorial cut of the atmosphere
showing the thermal structure as a function of pressure for several phases
during an orbit.
The rotation leads to a moving of the hottest zone with time.
As a consequence of the finite radiative timescale, the hottest zone is
shifted with respect to the maximum of insolation
(taking place at the substellar point).
Moreover, the eccentric orbit 
generates
a global decrease of temperature
from the first panel to the fourth as well as a uniformization of
the temperature with a lower day to night contrast.
During the periastron encounter, the high incoming stellar flux leads to a
thermal inversion (as shown in panels 1 and 5) as noted above.
Because of the finite radiative timescale, the inversion location is not
centered at the substellar point but is delayed in the direction of the rotation of about 10$^\circ$ (non-constant due to the speed difference between
the rotation and the heat propagation).

%\epsscale{0.5}

\paragraph{Flux ratio}

\begin{figure*}
\begin{center}
  \includegraphics[width=0.8\textwidth]{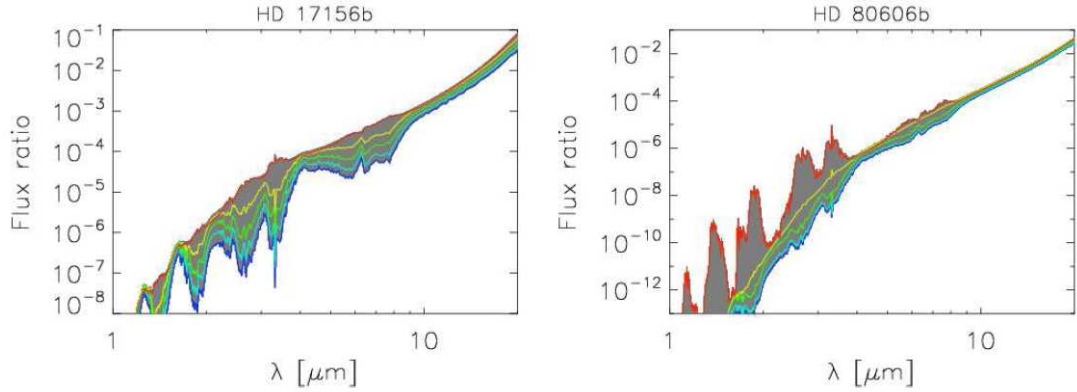}
\end{center}
  \caption{ Ratio of the planet flux by the star flux as a function of wavelength\label{fig:res2} for every times ({\it shaded area}) and at several positions ({\it lines}).\\
A color version of this figure is available in the online journal.
}
\end{figure*}

Figure~\ref{fig:res2} shows the planet to star flux ratio as a function of
wavelength.
For the coldest profiles, the features---in absorption---are mainly caused by
H$_2$O and CO.
As previously discussed in the models including a thermal inversion
\citep{Hubeny03,Burrows06,Fortney08},
the features are seen in emission near periastron due to the inversion of
temperature.

\paragraph{Predictions for \emph{Spitzer} observations}

Croll et al. have observed the system during the Cycle 5 of \emph{Spitzer} with IRAC
for 30 hr (pid: 50747).
The data have not been published yet.

\begin{figure*}
\begin{center}
  \includegraphics[width=0.8\textwidth]{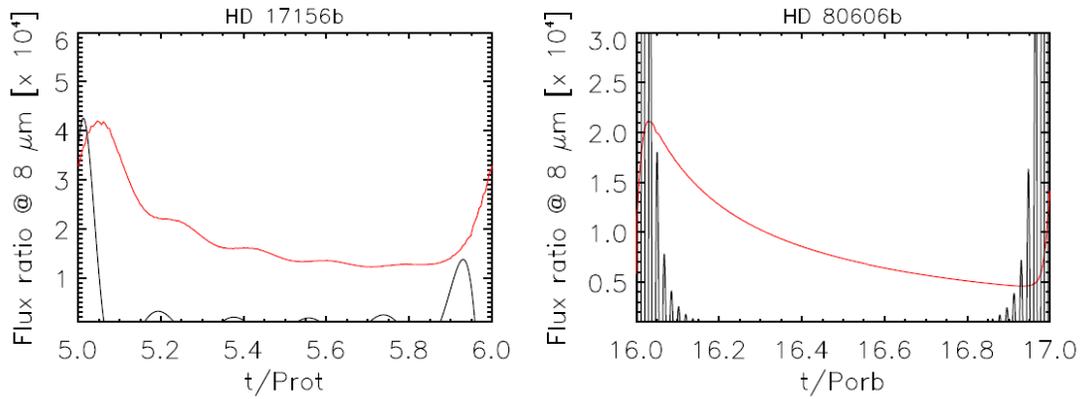}
\end{center}
  \caption{Evolution of the flux ratio ($\frac{F_{\rm plan.}}{F_\star}$)
in the {\it Spitzer} 8 $\mu$m as a function of time.\label{fig:resflux}
 We integrated the flux emited from the hemisphere seen from Earth; the viewing geometry from the orbit as well as due to the pseudo-synchronous rotation has been taken into account.
Note that the scale changes between the two panels.\\
A color version of this figure is available in the online journal.
}
\end{figure*}

Figure~\ref{fig:resflux} represents the predicted light curve in the
{\emph Spitzer} 8$\mu$m band.
It can be compared to Figure~4 of \citet{Irwin08}.
Our results have a smaller amplitude of variation, but we do not account
for winds.
The flux ratio in our model goes from $\sim 1.3 \times 10^{-4}$ to
$\sim 4.3 \times 10^{-4}$ during the orbit.

\subsubsection{HD~80606b}

\object{HD 80606b} has been discovered by \citet{Naef01}.
The high eccentricity of 0.932 
leads to a factor $\sim$~800 of insolation
between apastron and periastron.

\paragraph{Previous studies}

\citet{LL08} performed a hydrodynamic model of this planet.
They  predicted the planet to star ratio $F_{\rm p}/F_\star$ in the 8 $\mu$m
band to vary from
$\sim 4 \times 10^{-4}$
to $\sim 6.5 \times 10^{-4}$
from $\sim 20$ hr preceding periastron until $\sim 40$ hr following
periastron.

As recently published, \citet{Laughlin09} observed the secondary eclipse of this planet with
{\it Spitzer} in the 8~$\mu$m band.
The analysis of the light curve 
 is consistent with their predicted radiative time at the 570~mbar level 
of $\sim$~4.5hr.
\citet{Moutou09} confirmed that the planet transits its star and analyzed a transit event.
They inferred a planetary radius of $R_{\rm pl} = 0.86 \pm 0.1$~\rj .

\paragraph{Thermal structure}

The temperature contrasts from day side to night side are less pronounced than
in the case of HD~17156b,
due to a globally cooler planet and thus a longer radiative timescale as well
as a shorter rotation period
of $\sim$~1.72~days as we can see in Figure~\ref{fig:res1}.

Moreover, as we can see in Figure~\ref{fig:tempe}, the planet is globally more
isothermal below 10~mbar
than in the case of HD~17156b.
The temperature difference from 10~mbar to the highest pressures is less
than 100~K.
On the other hand, above 10~mbar, the temperature inversion---when present---is very large, reaching a 400~K peak.
This is due to the fact that although the planet is on average cooler
than HD~17156b, its periastron distance ($\sim$~0.029 AU, almost half
of HD~17156b's) is so short that during the periastron encounter the planet receives a blast of radiation from its star.

\begin{figure*}
\begin{center}
  \includegraphics[width=0.8\textwidth]{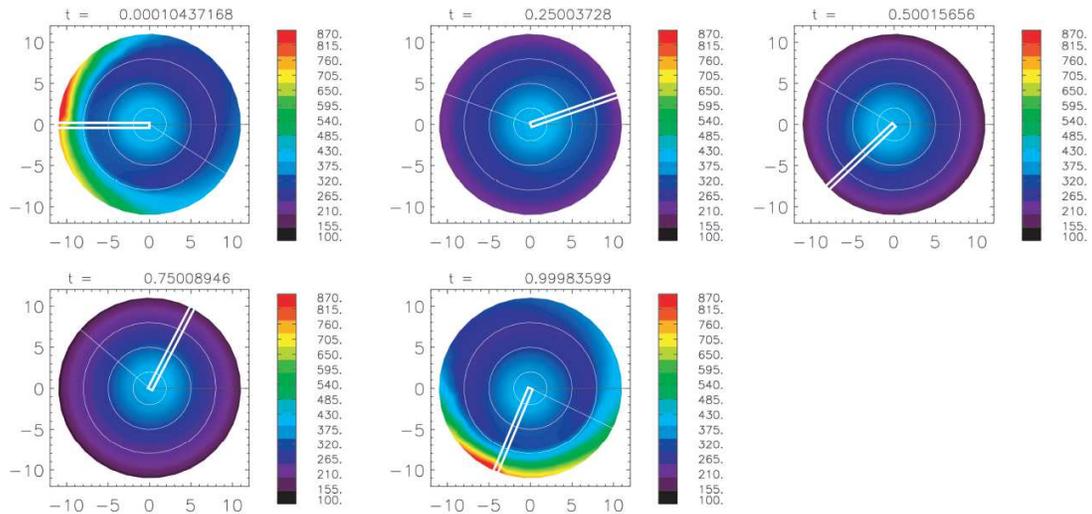}
\end{center}
  \caption{Same as Figure~\ref{fig:res_movie} for the planet HD80606b (with a different temperature scale).
    \label{fig:res_movie2}\\
An animation of this figure is available in the online journal.
  }
\end{figure*}

Figures~\ref{fig:res1} and \ref{fig:res_movie2} show that the very large
planet-star distance variation leads to a greater temperature difference
between periastron and apastron.
In Figure~\ref{fig:res_movie2}, we can see that the relatively fast rotation cause a 
large shift between the substellar point and the hottest spot---almost 180$^\circ$ in the first panel but highly varying since the planet
rotates faster than the temperature propagation
and also the radiative timescale depends on the atmosphere temperature.

%\epsscale{0.4}

Our consistent radiative transfer model however,
gives temperatures lower than presented by \citet{LL08}.
We reach temperatures up to 900~K, to be compared with the 1200~K obtained by
these authors.
In comparison, our model cannot represent latitudinal atmospheric flows but it has a consistent vertical structure.
Moreover, the long radiative timescale due to the cold temperatures of the planet prevent very large temperature contrasts to occur due to radiative effects alone.

In Figure~\ref{fig:res2}, we can see that the very important temperature
inversion leads to emission features that are very large.
In contrast, the cold temperatures of the distant apastron lead to
absorption features that are a lot fainter.

\paragraph{Predictions for \emph{Spitzer} observations}

In Figure~\ref{fig:resflux}, we can see that HD80606b's light curve is
smooth, due to the fast rotation of the planet, but contrary to the
hydrodynamic simulations of \citet{LL08}, we do not achieve a peak of
the flux ratio as strong as $8 \times 10^{-4}$ but rather $2.1 \times 10^{-4}$.

This small amplitude variation does not reproduce \citet{Laughlin09} observed light curve during secondary eclipse.
In this case, the very short but intense heating increase cannot generate the observed planetary flux if we account for radiation effects alone and we have to take into account hydrodynamic processes.

\section{Discussion}
\label{discuss}

\subsection{Role of the initial inputs to the model}

\begin{figure}[b!]
\begin{center}
  \includegraphics[width=0.4\textwidth]{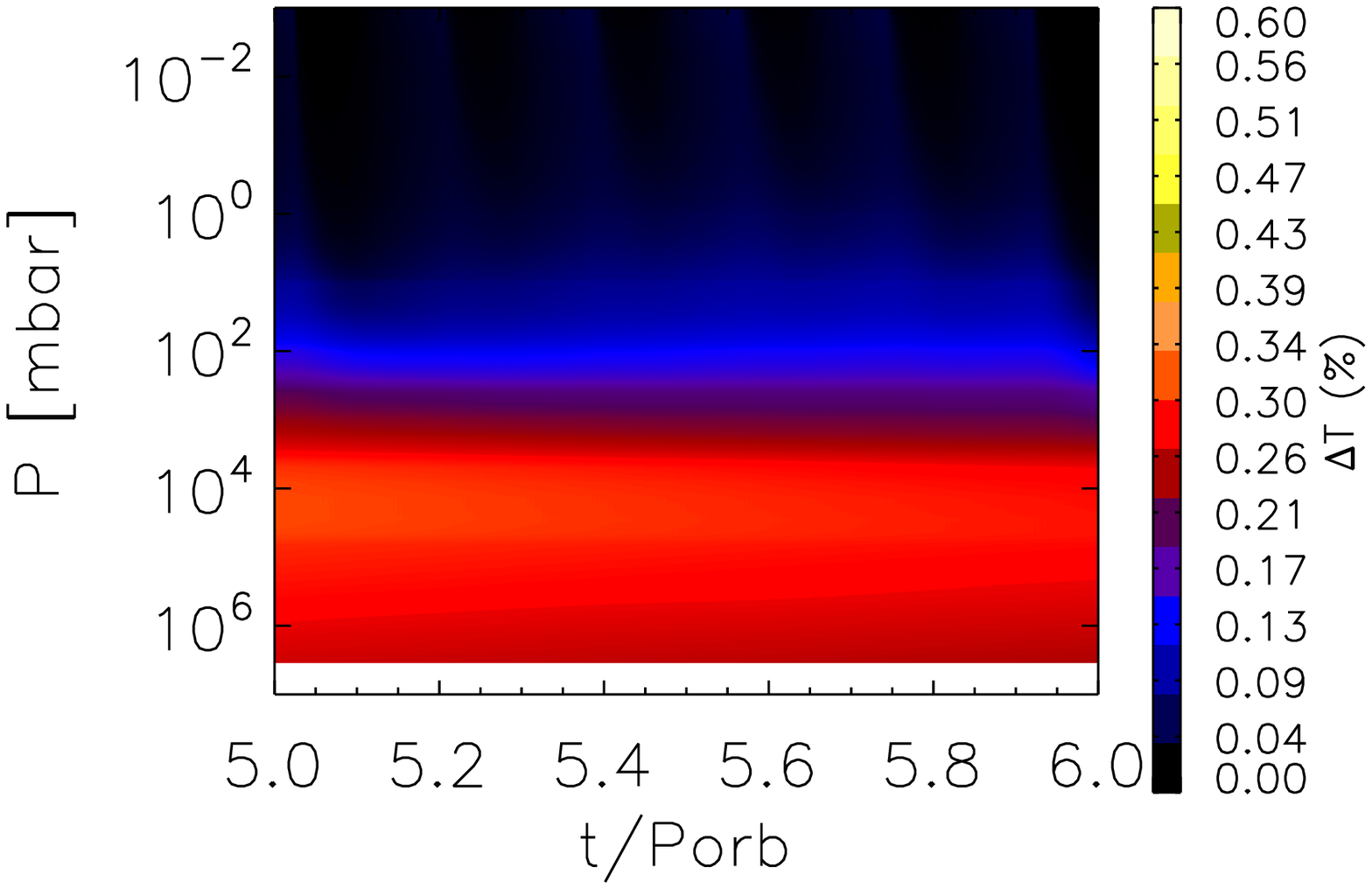}
\end{center}
  \caption{Relative difference of resulting temperature at each level between the two extremal initial profiles.\label{fig:dt}
Profiles using the initial conditions from the hottest and coldest profiles achieved by the nominal calculations for HD17156b have been calculated.
Here, the relative differences between the two extreme cases for each pressure level as a function of time are shown.\\
A color version of this figure is available in the online journal.
  }
\end{figure}

The main limitation of our model is that we do not vary the chemical
composition of the atmosphere in lock step with changes in the thermal
profile. In order to quantify the error produced by using a constant
abundance profile of the constituents, we conducted a test for the planet
HD~17156b assuming the two different extreme values for heating and
cooling rates, as well as the abundances, calculated for the hottest
and coldest profiles obtained by the previous study.
Although the calculations take a different amount of time before reaching
a periodic solution (the longer radiative timescale for a colder thermal
profile makes the model slower to converge), 
the final periodic solution gives similar results for a complete orbit.
Figure~\ref{fig:dt} shows the relative differences in the final thermal structure as a function of time using the two extremes initial temperature and heating/cooling rates.
The largest value is less than 0.5\%.

This is consistent with the estimation of the calculation time of
CO--CH$_4$
chemical reaction, which is the one determining the main spectral features of
the planet.
We calculated this reaction time using the prescription of \citet{Bezard02}.
For HD17156b, this time above the 250 mbar pressure level---where the temperature variations begin to be relatively important---is less than 10$^7$~s, more than the time of the orbit.
And the cooler planet HD80606b has even longer chemical timescales
%Indeed for the least favorable case in the (P,T) domain, this time is
$\sim 3 \times 10^{15}$~sec.
%\citep{Cooper06}, far longer than the time of the orbit,
These results justify our approach of not changing the chemical composition with time.

\begin{figure}
\begin{center}
  \includegraphics[width=0.4\textwidth]{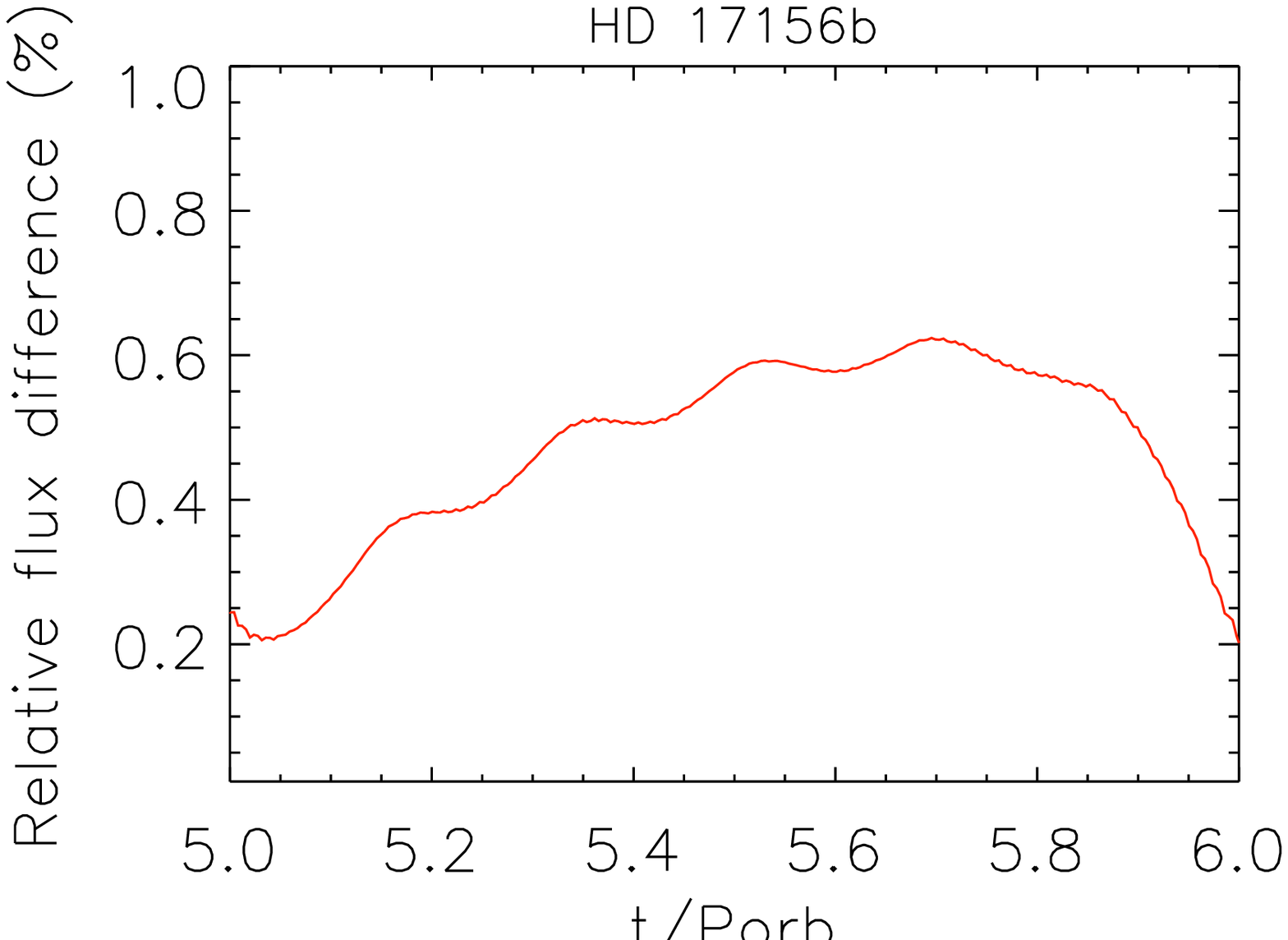}
\end{center}
  \caption{Relative difference of resulting flux ratio integrated in the 8~$\mu$m band.\label{fig:diff_flux}
As in Figure~\ref{fig:dt}, this shows that using the coldest or the hottest profile from the nominal calculations  for HD17156b as initial starting point has little effect on the final light curve obtained.
  }
\end{figure}

In order to confirm this conclusion, we calculated the flux viewed from Earth
integrated in the 8~$\mu$m band for the two extremal initial profiles for HD17156b.
This results in little differences, as shown in Figure~\ref{fig:diff_flux}.

We have adopted solar composition for all our calculations. If we increase the metallicity, we expect the general thermal structure to be slightly hotter due to the higher opacity and thus greater absorption of incoming flux \citep{Fortney06}. Global spectral features have been shown to exhibit only minor differences due to metallicity in the 8~$\mu$m spectral window \citep{Burrows06,Fortney09}. On the other hand, the radiative time scale is proportional to the atmospheric density \citep{Showman08} which increases with metallicity, but this timescale also decreases with temperature (higher metallicity atmospheres are hotter). These two effects  should compensate each other to a large extent. Consequently, we expect our qualitative conclusions to be insensitive to metallicity variations, to first order.

\subsection{Role of the rotation rate}
\label{discussrot}

It should be noted that the formalism of \cite{Hut81} contains uncertainties
in the calculation of the rotation period of the planet.
For instance, \cite{Ivanov07} using another formalism gave a rotation rate
equal to $\sim 1.5$ times
the circular orbit angular velocity at periastron
 $\Omega_{p}$ given by:
 $\Omega_{p}\sim \sqrt {GM_{\star}\over d_{\rm min}^{3}}$,
where $M_{\star}$ is the mass of the parent star.
With the parameters of HD~17156b, this gives a rotation period
of $P_{\rm spin}^{\rm IP}\sim 2.5$~days.

In order to test the effects of the period of rotation, 
we  used the nominal model for HD~17156b with twice the rotation rate 
and half the rotation rate given by Equation~\ref{eq:Hut81}
(resp. $P_{\rm spin}=\frac{1}{2}\times P_{\rm spin}^{\rm Hut}\sim 1.9$~days
and $P_{\rm spin}=2\times P_{\rm spin}^{\rm Hut}\sim7.6$~days) as well as
the rotation rate given by \cite{Ivanov07}
($P_{\rm spin}= P_{\rm spin}^{\rm IP}\sim 2.5$~days).

Moreover, a more rapid rotation rate can mimic a planet wherein
uniform zonal winds occur in addition to the rotating frame.

\begin{figure*}
\begin{center}
  \includegraphics[width=0.8\textwidth]{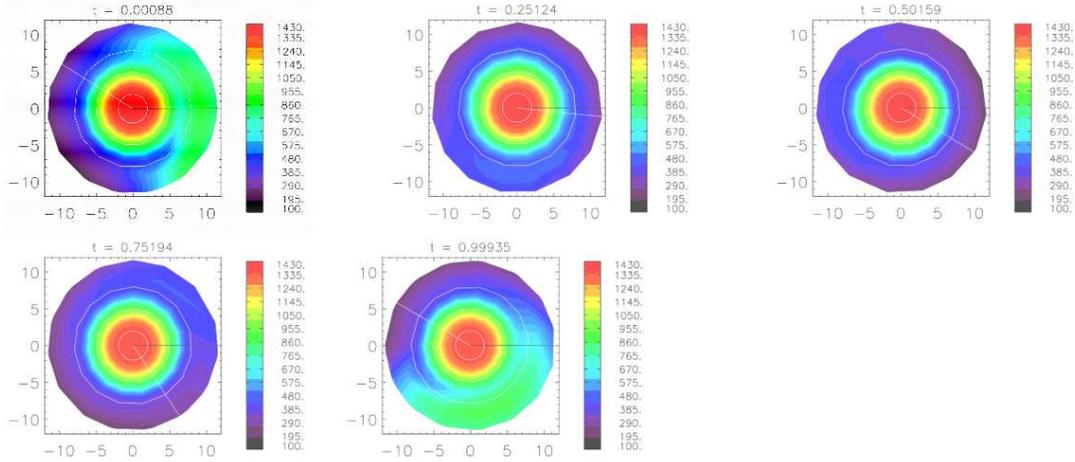}
\end{center}
  \caption{Same as Figure~\ref{fig:res_movie} for
    $P_{\rm  spin} = 2 \times P_{\rm  spin}^{\rm  Hut}$
    This figure is also available as an mpeg animation in the electronic.\label{fig:temp_2P}\\
An animation of this figure is available in the online journal.
  }
\end{figure*}

\begin{figure*}
\begin{center}
  \includegraphics[width=0.8\textwidth]{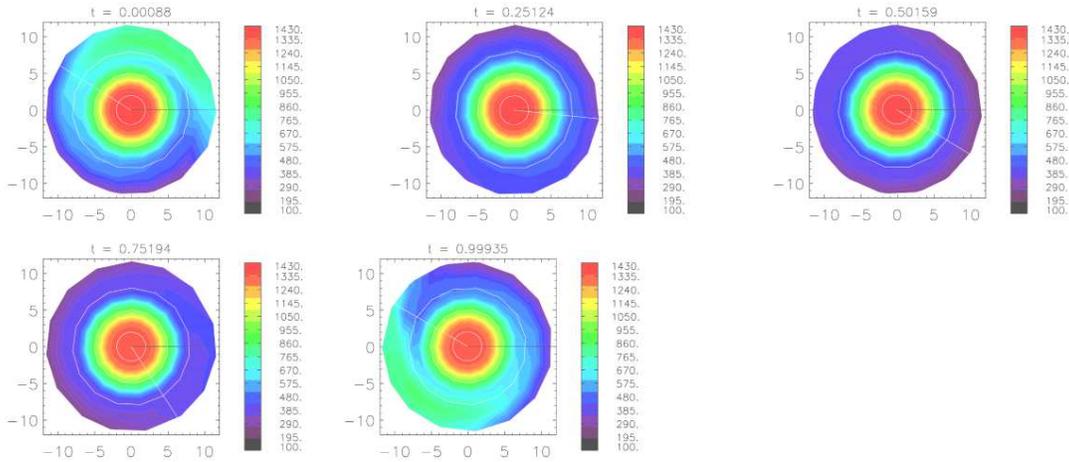}
\end{center}
  \caption{Same as Figure~\ref{fig:res_movie} for
    $P_{\rm  spin} = \frac{1}{2} \times P_{\rm  spin}^{\rm  Hut}$.\label{fig:temp_.5P}\\
An animation of this figure is available in the online journal.
  }
\end{figure*}

Figures~\ref{fig:temp_2P} and \ref{fig:temp_.5P} show the temperature
structure of the planet (equatorial cuts) as a function of time
for the two opposite cases.
The planet with a rotation rate 2 times faster than our nominal profile
is more uniform, as expected. 
The first and last panels of Figure~\ref{fig:temp_.5P} exhibit temperature
contrasts up to 300~K at 1~mbar and as the other panels
show there is almost no latitudinal contrast below this pressure level
when as the planet to star distance increases.
On the other hand,  with the slowest rotation rate as shown in
Figure~\ref{fig:temp_2P},
the day/night contrasts are more pronounced,
with a 500~K day/night contrast at the 1~mbar level near periastron and
still a $\sim 50$~K contrast far from periastron.

%\clearpage

\begin{figure*}
\begin{center}
  \includegraphics[width=0.7\textwidth]{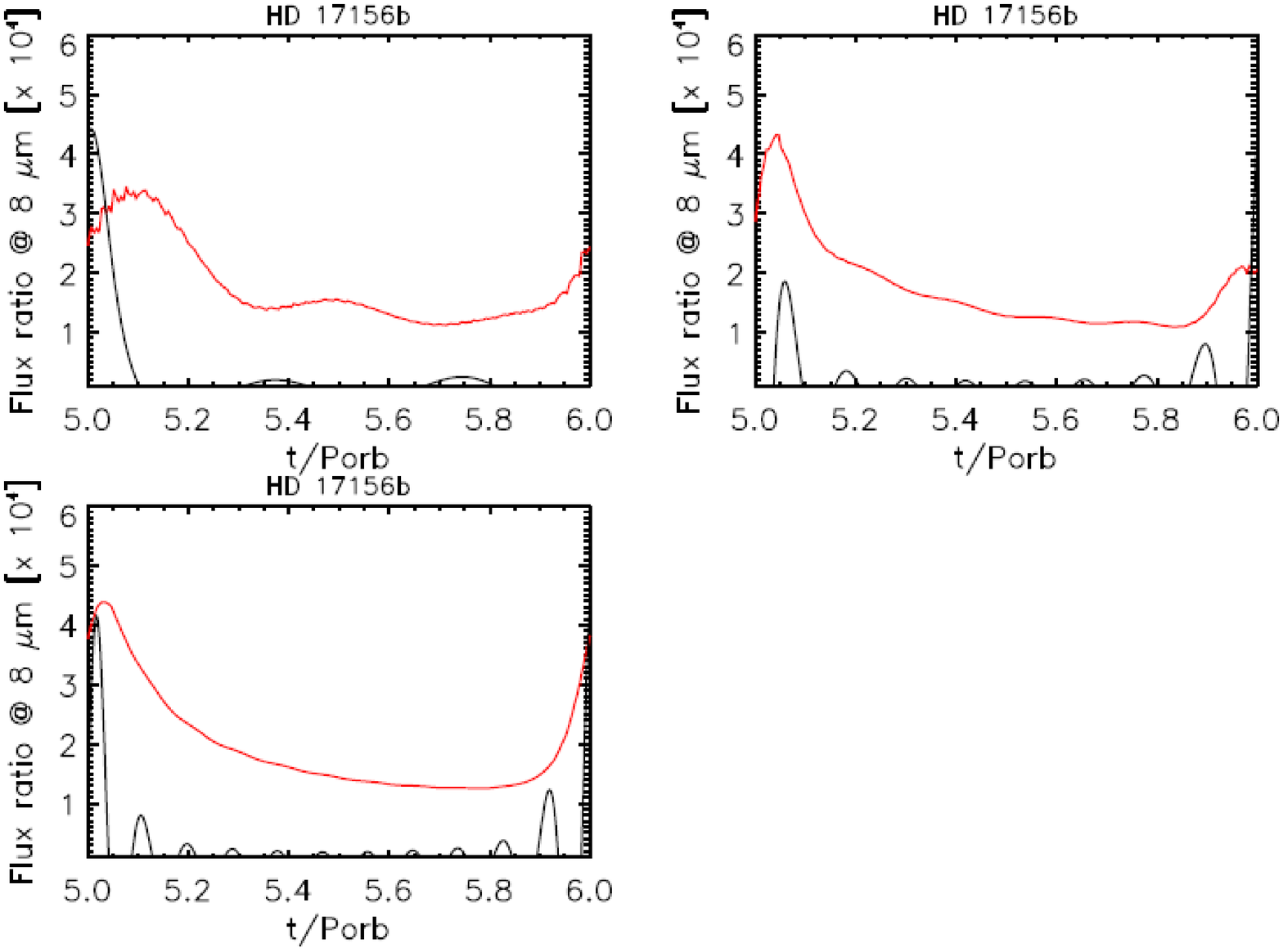}
\end{center}
  \caption{Same as Figure~\ref{fig:resflux} for
    $P_{\rm  spin} = 2 \times P_{\rm  spin}^{\rm  Hut}$ (first panel),
    $P_{\rm  spin} = P_{\rm  spin}^{\rm  IP}$  (second panel)
    and  $P_{\rm  spin} = \frac{1}{2} \times P_{\rm  spin}^{\rm  Hut}$
    (last panel)\label{fig:resflux2}.\\
A color version of this figure is available in the online journal.
}
\end{figure*}

Figure~\ref{fig:resflux2} shows the \emph{Spitzer} 8$\mu$m light curve for
the three cases.  The differences are large enough that observations
should be able to discriminate between the rotation rate predicted by
\citet{Hut81} or by \citet{Ivanov07}.

\section{Conclusions}

We presented a time-dependent radiative model for the atmosphere of the
transiting planets.
This model takes into account the eccentricity of their orbit as well as the
pseudo-synchronous rotation.

We applied our model to the planets HD~80606b and HD~17156b.  The
results define the radiative portion of the response of these
eccentric planets to time-variable stellar insolation and different
possible values of the their pseudo-synchronous rotation rate.

We showed that the high eccentricity makes the planets 
periodically transfer from a state where a stratospheric
temperature inversion
takes place to a cooler planet with monotonic temperature profiles
and vice versa.
Significant time variation in an eccentric
planet's spectral features would be expected based on our calculations,
and methods to observe such variations phased to the orbit would be very
valuable.

Moreover, the maximum of temperature and observable flux is delayed with respect to the maximum of stellar heating (periastrion), due to the finite radiative time

In some cases, an analysis of the planet during the periastron encounter is
feasible.
It will in principle allow us to discriminate which of the competing theory
that predict the pseudo-synchronous rotation period is correct
since the temporal variation of the planet flux is relatively highly
dependent on its rotation period.

Before reaching a time-dependent radiative model including hydrodynamics,
such observations will also help determine if the planet structure is mainly driven by radiation or
by atmospheric dynamics. 

In the case of HD80606b, the heating increase takes place for a such short time that the radiative model fails to explain the observed flux variation without accounting for hydrodynamic effects.

\acknowledgments

This research was supported by an appointment to the NASA Postdoctoral Program
at the Goddard Space Flight Center, administrated by Oak Ridge Associated
Universities through a contract with NASA.

The calculations presented in this paper have been performed with
\emph{Columbia}, a High End Computing NASA facility through the
Science Mission Directorate Program
and with the SIO's \emph{Quadri} cluster at the Observatoire de Meudon.

\end{document}